\documentclass[preprint,12pt]{elsarticle}

\usepackage{amssymb}
\usepackage{bm}
\usepackage[hidelinks]{hyperref}
\usepackage{float}
\usepackage{amsmath}

\journal{Computers in Biology and Medicine}

\begin{document}

\begin{frontmatter}

\title{A CT-based deep learning system for automatic assessment of aortic root morphology for TAVI planning }

\author[poli]{Simone Saitta}
\ead{simone.saitta@polimi.it}
\author[psd1,poli]{Francesco Sturla}
\author[psd2]{Riccardo Gorla}
\author[psd2]{Omar A. Oliva}
\author[poli]{Emiliano Votta}
\author[psd2]{Francesco Bedogni}
\author[poli]{Alberto Redaelli}

\affiliation[poli]{organization={Department of Information, Electronics and Bioengineering, Politecnico di Milano},
            city={Milan},
            country={Italy}}

\affiliation[psd1]{organization={3D and Computer Simulation Laboratory, IRCCS Policlinico San Donato},
            city={San Donato Milanese},
            country={Italy}}
            
\affiliation[psd2]{organization={Department of Clinical and Interventional Cardiology, IRCCS Policlinico San Donato},
            city={San Donato Milanese},
            country={Italy}}

\begin{abstract}
Accurate planning of transcatheter aortic implantation (TAVI) is important to minimize complications, and it requires anatomic evaluation of the aortic root (AR), commonly done through 3D computed tomography (CT) image analysis. Currently, there is no standard automated solution for this process. 
Two convolutional neural networks (CNNs) with 3D U-Net architectures (model 1 and model 2) were trained on 310 CT scans for AR analysis. Model 1 performed AR segmentation and model 2 identified the aortic annulus and sinotubular junction (STJ) contours. Results were validated against manual measurements of 178 TAVI candidates. After training, the two models were integrated into a fully automated pipeline for geometric analysis of the AR.
The trained CNNs effectively segmented the AR, annulus and STJ, resulting in mean Dice scores of 0.93 for the AR, and mean surface distances of 1.16 mm and 1.30 mm for the annulus and STJ, respectively. Automatic measurements were in good agreement with manual annotations, yielding annulus diameters that differed by 0.52 [-2.96, 4.00] mm (bias and 95\% limits of agreement for manual minus algorithm). Evaluating the area-derived diameter, bias and limits of agreement were 0.07 [-0.25, 0.39] mm. STJ and sinuses diameters computed by the automatic method yielded differences of 0.16 [-2.03, 2.34] and 0.1 [-2.93, 3.13] mm, respectively. 
The proposed tool is a fully automatic solution to quantify morphological biomarkers for pre-TAVI planning. The method was validated against manual annotation from clinical experts and showed to be quick and effective in assessing AR anatomy, with potential for time and cost savings.	
\end{abstract}


\end{frontmatter}

\section{Introduction}
Transcatheter aortic valve implantation (TAVI) has emerged as an alternative to traditional open-heart surgery to treat severe aortic stenosis, proving effective in reducing morbidity and mortality in high-risk patients \cite{leon2016transcatheter, mack20155}. Despite its benefits, TAVI still carries risk for post-operative complications and drawbacks, including paravalvular leakage, device migration, annulus rupture and conductive disturbances \cite{bhushan2022paravalvular, breitbart2021implantation, khosravi2018tavi}. A meticulous preprocedural planning is thus crucial for minimizing the risk of complications and should envisage an accurate anatomic assessment of the aortic root apparatus, which is essential for selecting the optimal prosthetic device size \cite{astudillo2020automatic}. Accurate preoperative assessment is centered around the aortic annulus, and includes quantification of its diameters \cite{cerillo2012sizing}, angulation \cite{gorla2021impact} and perimeter \cite{schultz2010three}. Furthermore, a comprehensive analysis of the whole aortic root can provide anatomical measurements of the sinotubular junction (STJ) and sinuses of Valsalva \cite{elattar2016automatic, gorla2021impact, queiros2017automatic}. To this purpose, three-dimensional (3D) computed tomography (CT) angiography is the preferred imaging modality to quantify aortic root (AR) anatomy before a TAVI \cite{delgado2011automated}. A comprehensive CT-based TAVI planning involves three main operations: segmentation of the anatomy, landmark detection, and measurement extraction \cite{astudillo2020automatic, elattar2016automatic, lalys2019automatic, tahoces2021deep, zheng2010automatic}. However, there is currently no standardized fully automated solution, and taking the necessary measurements can be a time-consuming process which often involves several manual operations that introduce operator-dependency and may limit reproducibility. Automatic and semi-automatic systems have already been proposed to identify and quantitatively assess AR features from CT images. Lalys et  al. \cite{lalys2019automatic} exploited semi-automatic segmentation tools relying on atlas-based methods to segment the AR, localize a wide range of anatomical landmarks (e.g., leaflet and coronary ostium positions) and obtain accurate quantification of annulus diameter. More recent approaches have taken advantage of convolutional neural networks (CNNs) to fully automate the detection of AR landmarks from 3D CT \cite{astudillo2020automatic, tahoces2021deep}. Given the black-box nature of deep neural networks, when employing these tools extensive validation against manual landmark tracing by experts should be performed. To date, the largest patient validation set has been reported by Astudillo et al. \cite{astudillo2020automatic}, who proved the feasibility of building fast and accurate CNN-based systems for detection of the three aortic cusp nadirs and coronary ostia, training their model on 444 CT scans and validating their landmark detection accuracy on 100 patients. Nonetheless, the proposed approach could only be applied to contrast-enhanced CT images and did not include automatic measurement extraction. Hence, full automation of the entire process, including segmentation, landmark detection and extraction of aortic features relevant for TAVI, is still lacking. Given the increasing adoption of TAVI also in intermediate and low-risk patients \cite{leon2016transcatheter}, a reliable, fast and efficient  method for assessing AR anatomy and determining the appropriate device size could have an increasingly broad impact by making TAVI plannig faster, reliable and fully repeatible.
In this work, we sought to combine deep learning techniques and tools from differential geometry to build a fully automatic system that segments the AR from 3D CT data, extracts AR-specific anatomical landmarks and computes clinically relevant measurements for TAVI planning. The effectiveness of the proposed method was extensively evaluated on a group of 178 patients.

\section{Methods}

\subsection{Data collection and manual annotation}
CT scans of 512 subjects acquired between 2010 and 2022 were retrospectively collected. Pixel spacing ranged from 0.26 x 0.26 mm2 to 0.87 x 0.87 mm2, while slice thickness ranged between 0.25 and 1 mm. 24 acquisitions were excluded because of reconstruction artifacts, presence of metal devices, or with slice thickness greater than 1 mm. Accordingly, the dataset was split in two subsets: dataset A, for which no AR measurement was available (N=310), and dataset B (N=178), for which an expert operator manually took measurements of the aortic annulus, STJ and Valsalva sinuses plane using the commercial software 3mensio Structural Heart (v8.2, Pie Medical Imaging, Maastricht, Netherlands) (Figure \ref{fig1}). After manual positioning, 3mensio allowed to compute maximum and minimum diameters, area, and perimeter of each plane. All patients in dataset B were TAVI candidates with severe aortic stenosis.
For all CT scans in dataset A, an initial segmentation of the ascending aorta was obtained using a neural network previously trained by our group \cite{saitta2022deep}. Segmentations that were erroneously inferred (e.g. incomplete filling) by the pre-trained model were adjusted by an experienced operator using a semi-automatic region growing algorithm \cite{yushkevich2016itk}. All the resulting aortic segmentations include the left ventricle outflow tract (LVOT), the whole aortic root and the proximal portion of the ascending aorta according to the scan-specific field-of-view. Both aortic annulus and STJ were manually segmented for all CT scans in dataset A. To expedite the annotation process, a graphic-user-interface (GUI) was appositely developed using VTK \cite{schroeder2000visualizing}. Through the designed GUI, the user could easily position a plane in the 3D image space and, exploiting the previously obtained segmentation of the aorta, the annulus or STJ labels were automatically assigned to the image points within an Euclidean distance of 3 mm the chosen plane and inside the aorta.
All image data used in the present study were collected from IRCCS Policlinico San Donato (Milan, Italy). The study was approved by the local ethics committee and informed consent was waived because of the retrospective nature of the study and the analysis of anonymized data.

\subsection{Neural network training}
\begin{figure}[H]
	\centering
	\includegraphics[width=0.6\textwidth]{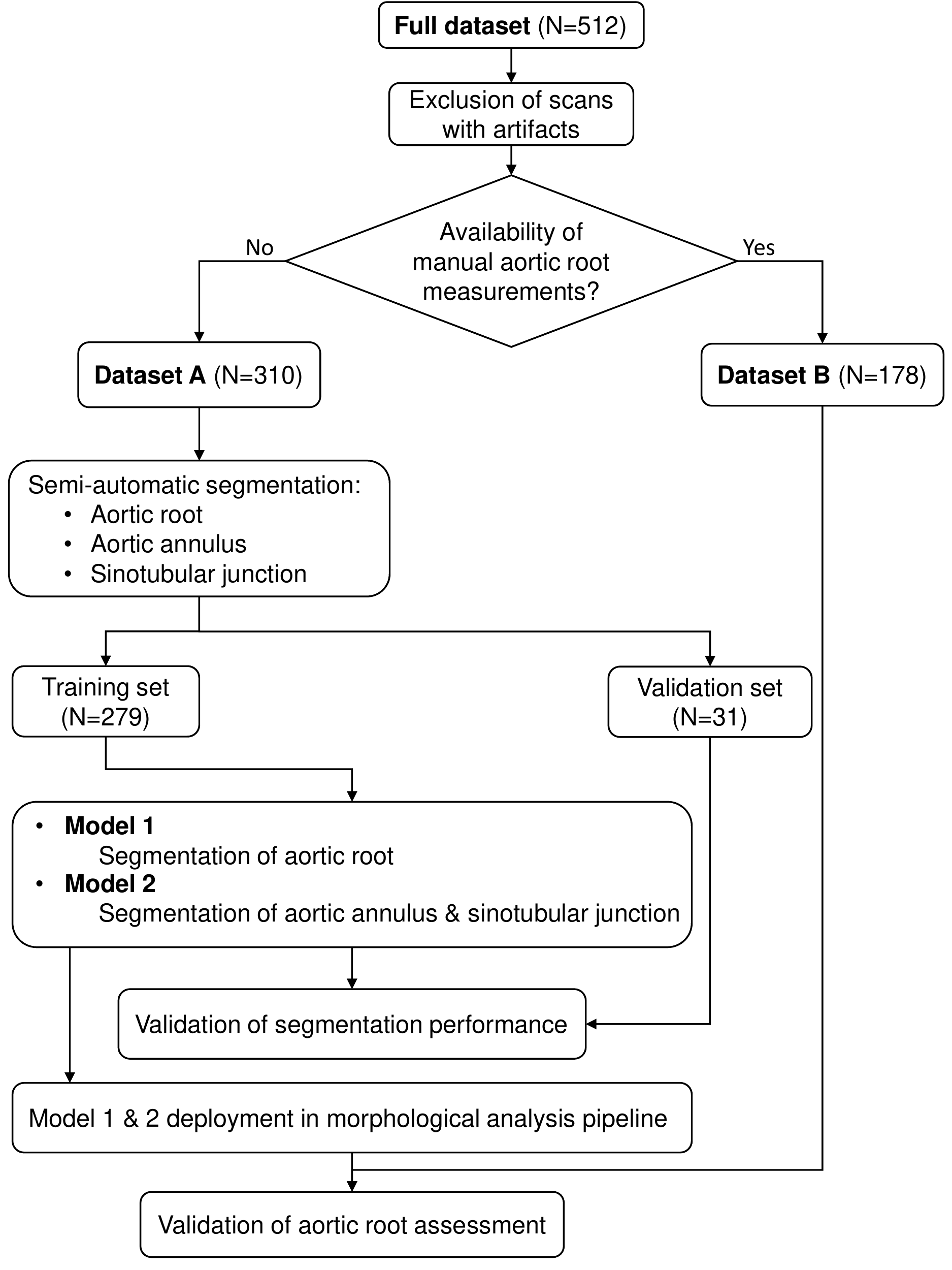}
	\caption{Schematic representation of the adopted workflow. Dataset A  includeds manual segmentations of the aortic root, annulus and sinotubular junction (STJ); it was used to train and validate two neural networks (model 1 and model 2) for automatic segmentation. Dataset B included manual annotations of aortic root measurements, used to validate the developed morphological analysis pipeline.}
	\label{fig1}
\end{figure}

Dataset A was divided into training and test sets: 279 (90\%) scans with their corresponding ground truth segmentations of the aorta, annulus and STJ were randomly selected and used to train two different neural networks, respectlively dedicated to the segmentation of the aorta (model 1) and to the multi-class segmentation of aortic annulus and STJ (model 2). The remaining 31 (10\%) scans were used for validation of both models (Figure \ref{fig1}). Both models were based on a fully convolutional 3D U-Net architecture proposed in \cite{isensee2021nnu}, with encoding and decoding branches of 5 resolution levels each, defined using residual units as introduced by \cite{zhang2018road}. Parametric rectified linear unit (PReLU) activation functions \cite{he2015delving} were used at each residual layer. A data augmentation routine including random Gaussian noise, cropping, mirroring and rotation was implemented in the MONAI framework \cite{cardoso2022monai}. A Dice loss was used to train model 1, while a combination of Dice and Focal loss \cite{yeung2022unified} with equal weights was used to train model 2. Training was carried out on an NVIDIA A100 GPU over 1000 epochs, using an Adam optimizer with learning rate of 0.0001.

\subsection{Pipeline implementation}
\begin{figure}[H]
	\centering
	\includegraphics[width=0.7\textwidth]{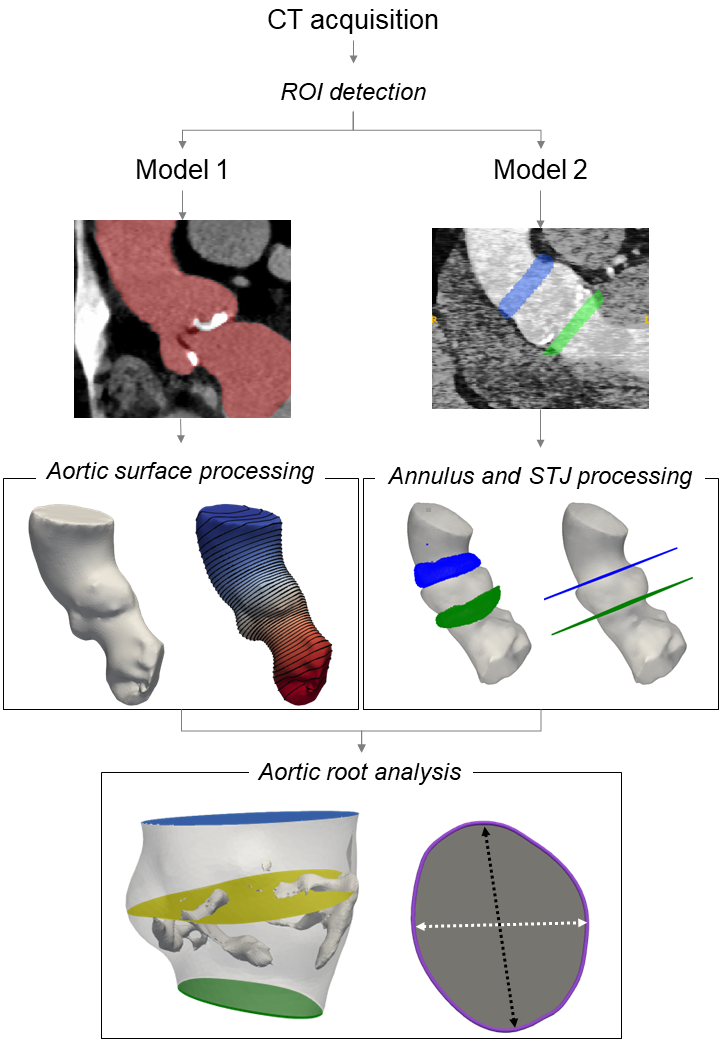}
	\caption{Schematic representation of the implemented automatic pipeline. The ROI is detected from the input CT scan. Model 1 infers the segmentation of the aortic root with the left ventricle outflow tract (LVOT), shown as the red label (left side). Model 2 infers the segmentation of the aortic annulus and STJ (in green and blue, respectively) (right side). The aortic surface processing step computes the second eigenvector of the LBO and its contours (left side). The annulus and STJ processing step performs refinement of model 2 segmentation. In the aortic root analysis step (bottom), the aortic root is isolated, and the Sinuses of Valsalva are detected. Anatomical measurements are computed together with calcium volume (shown in white, bottom panel).}
	\label{fig2}
\end{figure}

After training, models 1 and 2 were embedded in a fully automated pipeline for aortic root analysis and TAVI pre-procedural planning. The different sequential steps of the implemented pipeline are described below and exemplified in Figure \ref{fig2}.

\subsubsection{Aortic root region detection}
\begin{figure}[H]
	\centering
	\includegraphics[width=\textwidth]{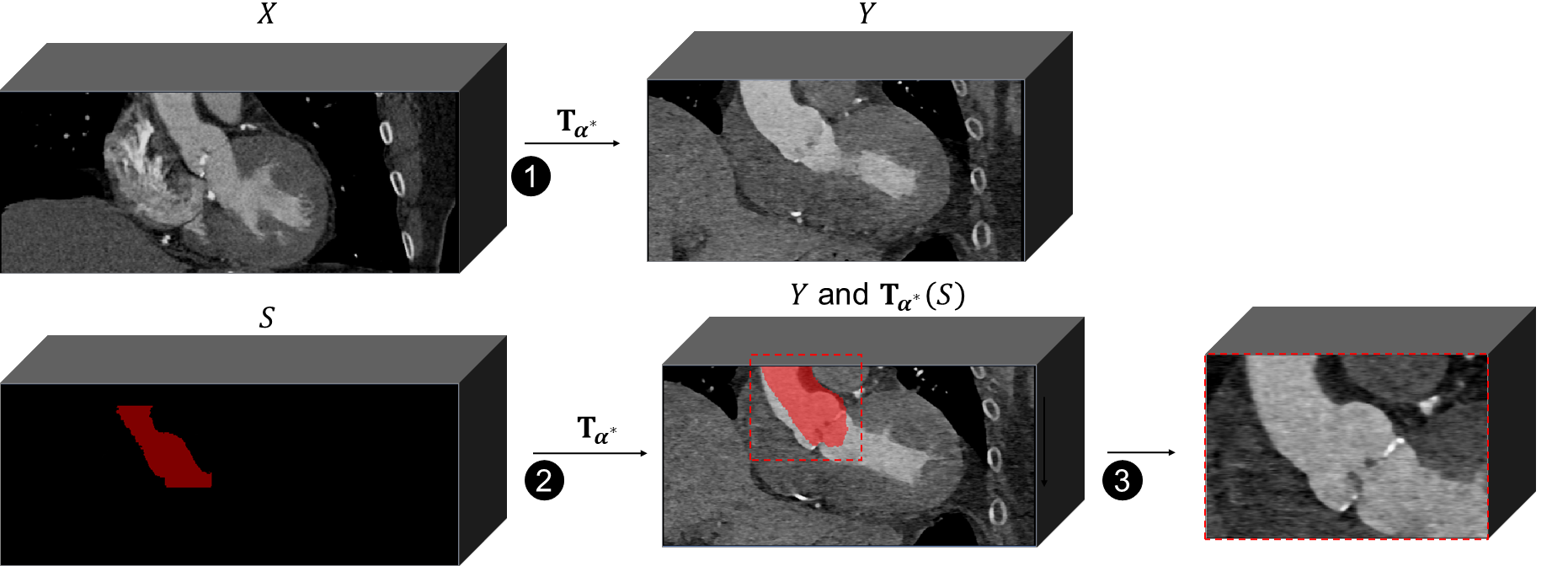}
	\caption{Template-matching approach for automatic identification of the region of interest (ROI) from 3D computed tomography (CT). A template image ($X$) for which aortic segmentation ($S$) is known, is affinely registered to the processed image ($Y$). The transformation mapping $X$ to $Y$ ($\mathbf{T}_{\bm{\alpha}^*}$) is applied to $S$ and the bounding box of $\mathbf{T}_{\bm{\alpha}^*}(S)$ is used to crop the ROI from $Y$.}
	\label{fig3}
\end{figure}

The first step of the pipeline consists of cropping out from the 3D CT scan a region of interest (ROI) encompassing the aortic root (Figure \ref{fig3}). Automatic ROI detection is achieved through a 3D template-matching approach \cite{marstal2016simpleelastix}. An image X with its ground truth aortic root segmentation S is selected from dataset A. When processing a new case (target image, $Y$), the image $X$ is roughly aligned to the target image using a rigid transformation ($\mathbf{R}$) followed by an affine transformation ($\mathbf{A}$) able to account for scaling and shear deformations. Defining the generic final transformation: 
\begin{equation}
	\mathbf{T}_{\bm{\alpha}^*}(X) = (\mathbf{A} \circ \mathbf{R})(X),
\end{equation}

parametrized by $\bm{\alpha}^*$, the optimal mapping is found as:

\begin{equation}
	\bm{\alpha}^* = arg\max_{\bm{\alpha}} I(X, Y),
\end{equation}

where $I(X, Y)$ is the mutual information between the two images, which can be considered a nonlinear generalization of cross-correlation \cite{maes2003medical}. $\bm{\alpha}^*$ is found using a gradient descent algorithm. Once found, $\mathbf{T}_{\bm{\alpha}^*}$ is used to map the known ground truth aortic root segmentation into the target image domain, obtaining $\mathbf{T}_{\bm{\alpha}^*}(S)$. The bounding box of $\mathbf{T}_{\bm{\alpha}^*}(S)$ is dilated outward in each direction by 20 pixels and the resulting ROI is cropped out from the original image.

\subsubsection{Aortic surface processing}
After ROI cropping, the CT scan is processed by model 1, generating a binary mask of the ascending aorta including the LVOT. A marching cubes algorithm extracts the corresponding contour as a triangulated surface; surface smoothing is applied with a windowed sinc function interpolation kernel with passband of 0.01, and adaptive remeshing is performed following the approach described in \cite{valette2008generic}. For each processed surface, the Laplace-Beltrami operator (LBO) \cite{reuter2006laplace} is computed. LBO eigenvectors form a set of bases for the definition of continuous functions on the surface manifold. For tubular structures with a dominant longitudinal direction, the second LBO eigenvector (E1) represents a scalar field approximating the curvilinear abscissa of the surface centerline; E1 isocontours on the aortic surface are shown in Figure \ref{fig2}. 

\subsubsection{Annulus and STJ processing}
\begin{figure}[H]
	\centering
	\includegraphics[width=\textwidth]{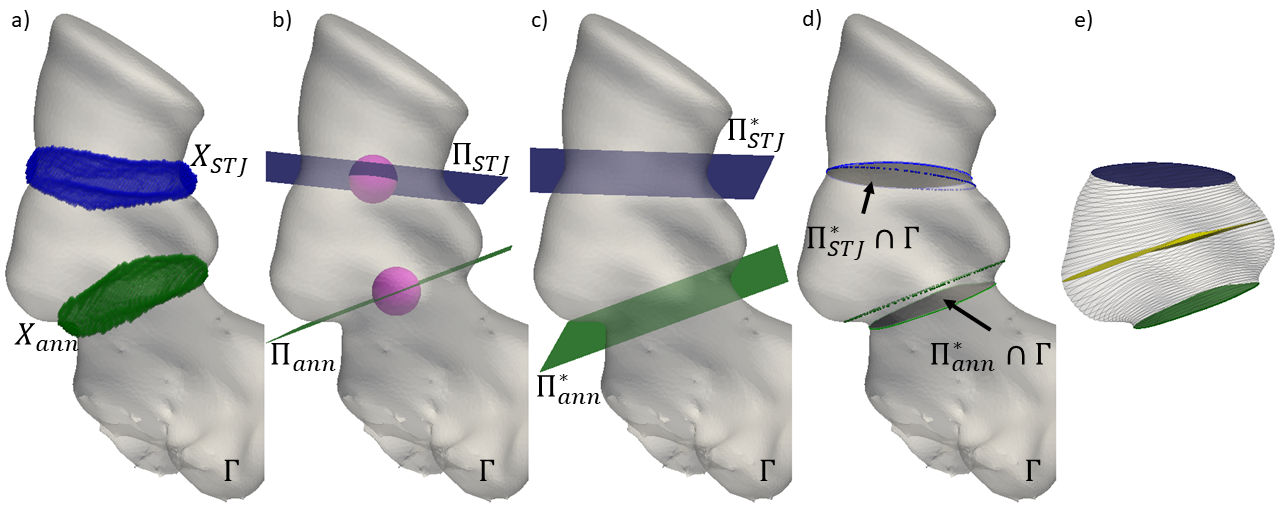}
	\caption{steps implemented for aortic annulus and STJ plane detection refinement. a) the point sets yielded by model 2 for the annulus (green) and STJ (blue), visualized with the aortic segmentation produced by model 1 (transparent gray). b) the best fitting planes for the two structures ($\Pi_{ann}$ and $\Pi_{STJ}$), together with the region inside which their center of mass is allowed to move during the refinement procedure (magenta spheres). c) the resulting planes ($\Pi_{ann}^*$ and $\Pi_{STJ}^*$) after the optimization procedure. d) visualization of the refined planes in dark gray and the initialization planes obtained by model 2. e) the isolated aortic root with the contours of the second eigenvalue of the Laplace-Beltrami operator (LBO) (in black) and the plane identifying the sinuses of Valsalva (yellow).}
	\label{fig4}
\end{figure}

The cropped CT scan is processed by model 2, which infers the masks and the corresponding set of points $X_{ann}$ and $X_{STJ}$ for the annulus and STJ, respectively (Figure \ref{fig4}). Using a random sample consesus (RANSAC) iterative algorithm with inlier threshold of 1.5 mm, two best-fitting planes are identified, namely $\Pi_{ann}$ and $\Pi_{STJ}$ from $X_{ann}$ and $X_{STJ}$, respectively. Specifically, a generic plane $\Pi$ is represented by a bounded rectangular mesh grid identified by its normal vector $\bm{n}$ and center of mass $\bm{c}$ : $\Pi = \Pi (\bm{n}, \bm{c})$. The normal vector and center of mass of both $\Pi_{ann}$ and $\Pi_{STJ}$, are refined (rotated and shifted) following a constrained optimization procedure, initialized using $\bm{n}$  and $\bm{c}$  computed by RANSAC. The optimal plane normal vector $\bm{n}^*$ and center of mass $\bm{c}+\bm{\delta}$ are defined as the ones that minimize the area of intersection with the aortic surface:

\begin{equation}\label{eq:planeOpt}
	\begin{array}{c}
	\Pi^* = \min\limits_{\bm{n}^*, \bm{\delta}} \: area(\Pi(\bm{n}^*, \bm{c}+\bm{\delta}) \cap \Gamma), \\
	
	\text{subject to} \: \lVert \bm{\delta} \rVert < K,
	\end{array}
\end{equation}
where $\Gamma$ represents the aortic root and LVOT surface and $K=5mm$. The described procedure is applied to both $\Pi_{ann}$ and $\Pi_{STJ}$ and the yielded planes are used to clip and isolate the AR (Figure \ref{fig4}.d). Within the extracted AR surface, the E1 isocontour with maximum area is used to identify the plane of the sinuses of Valsalva (Figure \ref{fig4}e).

\subsubsection{Aortic root analysis}
For the annulus, the STJ and the plane of Valsalva sinuses, the following metrics are automatically computed: i) area; ii) perimeter; iii) the maximum ($D_{max}$), iv) minimum ($D_{min}$) and v) mean ($D_{mean}=(D_{max}+D_{min}) / 2$) diameters. Maximum and minimum diameters are defined as the lengths of the largest and shortest segments connecting two opposite points of the perimeter while passing through the center of mass. In addition, for the aortic annulus the annulus angle, i.e., the angle formed by the plane normal to the foot-head image axis \cite{gorla2021impact}, is extracted. The AR calcium score  is automatically quantified as the volume enclosing pixels with Hounsfield units (HUs) greater than 800 by a simple thresholding of the corresponding  segmented AR \cite{bettinger2017practical}. 

\subsection{Statistical analysis}
Measurements based on expert manual annotations were used as reference values in assessing the proposed method's accuracy. Accuracy of areas, perimeters and diameters was evaluated using Bland-Altman analysis. Analyses were performed using python scipy 1.5.1 statistics library. 

\section{Results}
\subsection{Evaluation of segmentation performance}
\begin{figure}[h]
	\centering
	\includegraphics[width=\textwidth]{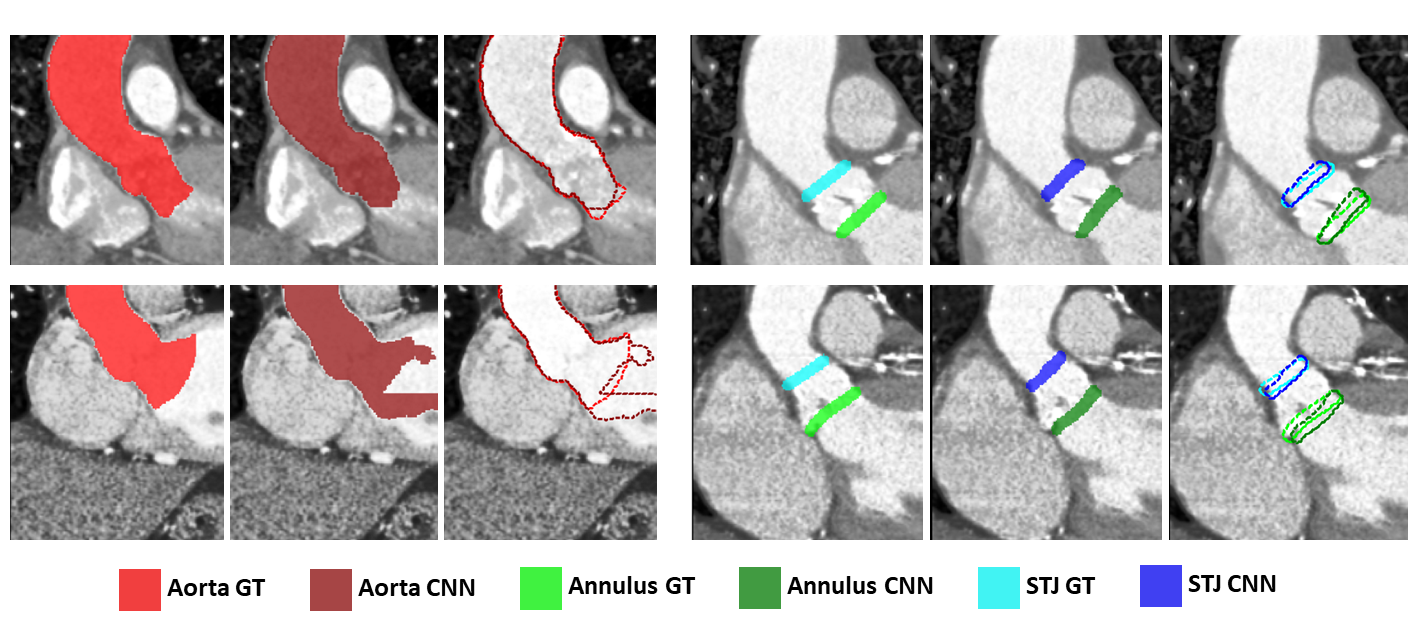}
	\caption{Comparison of manual (ground truth, GT) segmentation in bright red (aorta), bright green (annulus) and light blue (STJ) vs. automatic segmentation obtained by the trained neural networks (CNN) in dark red (aorta), dark green (annulus) and dark blue (STJ) on four test cases: the best (top row) and the worst (bottom row) for the two models. Coronal slices with color fill and with dashed contours are shown.}
	\label{fig5}
\end{figure}

The generalization performance of the trained models was evaluated on the validation set (n=31), for which ground truth labels of the aorta, annulus and STJ were available (dataset A). For the aortic segmentation, Dice scores and mean surface distances (MSDs) were computed for the aortic root, bounded by the aortic annulus and STJ planes. For all subjects in the validation set, both models 1 and 2 proved able to accurately trace the region encompassing the aortic root, annulus and STJ. Segmentation performance was quantitatively assessed, in terms of Dice coefficient and MSD with respect to manual ground truth segmentations; for both indices, mean value and [min, max] range were computed. For the aortic root, Dice score and MSD were equal to 0.93 [0.82, 0.98] and 1.10 [0.22, 5.99] mm, respectively. For the aortic annulus, mean Dice score and MSD were equal to 0.68 [0.34, 0.23] and 1.16 [0.90, 1.29] mm; for the STJ, mean Dice score of 0.70 [0.69, 0.21] and MSD of 1.30  [1.02, 8.49] mm were obtained. The test cases corresponding to the best and worst performance are shown in Figure \ref{fig5}.

\subsection{Comparison vs. manual measurements}
\begin{figure}[H]
	\centering
	\includegraphics[width=0.9\textwidth]{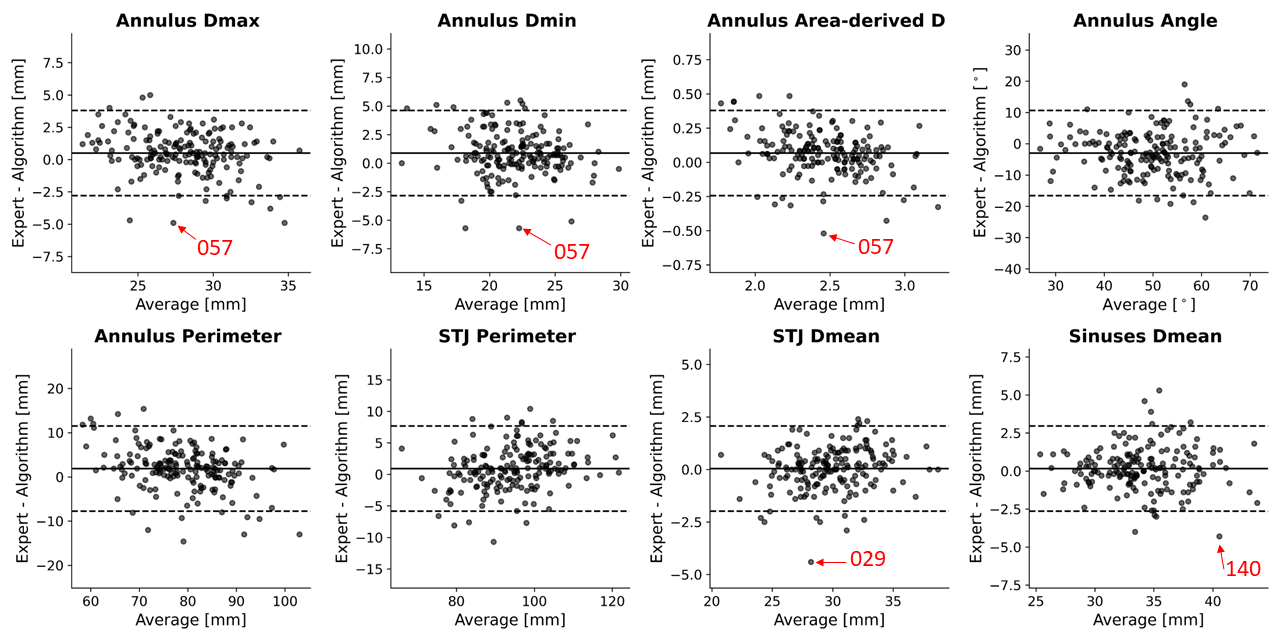}
	\caption{Bland–Altman plots of the proposed algorithm versus expert manual measurements. Mean differences are shown as continuous horizontal lines, while 95\% limits of agreement are shown as dashed horizontal lines. Critical cases are highlighted with a red arrow and identification number. }
	\label{fig6}
\end{figure}

Bland-Altman plots (Figure \ref{fig6}) allowed to analyze the agreement between the automatic pipeline output and manual measurements. Herein, we report comparisons as biases and 95\% limits of agreement, i.e., bias [lower limit, upper limit], where differences were computed as manual – algorithm. In general, automatic and manual anatomical measurements were in good agreement. A tendency of the automatic system to underestimate annulus diameters vs. manual measurements was observed, as evidenced by positive biases (solid horizontal lines in Figure \ref{fig6}). For $D_{max}$, measurement differences were 0.51 [-2.79, 3.81] mm. A similar trend was found for $D_{min}$, resulting in 0.89 [-2.8, 4.62] mm. Significantly smaller discrepancies between the two measurement techniques were obtained for the annulus area. Evaluating the area-derived diameter, bias and limits of agreement were 0.07 [-0.24, 0.38] mm. For the annulus angle, an average difference \textless 3\textdegree was found between measurements, while limits of agreement were [-17\textdegree, 11\textdegree]. $D_{mean}$ computed at the STJ and Sinuses by the automatic method slightly underestimated manual measurements, yielding differences of 0.05 [-1.98, 2.07] and 0.17 [-2.63, 2.97] mm, respectively. As compared to diameter measurements, perimeter measurements showed larger differences: -1.8 [-8.06, 11.74] mm for the annulus perimeter, and 1.09 [-6.18, 8.37] mm for the STJ perimeter.

\subsubsection{Critical cases}
\begin{figure}[H]
	\centering
	\includegraphics[width=0.5\textwidth]{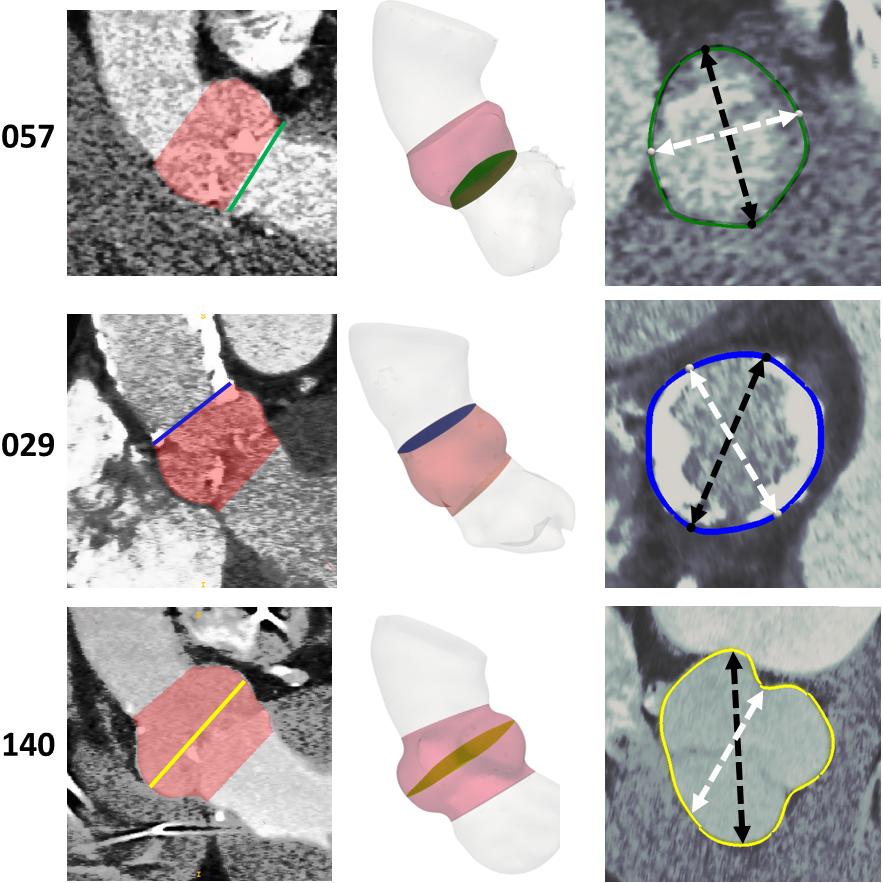}
	\caption{Results of the implemented automatic pipeline for 3 cases for which the discrepancies between expert and automatic measurements were largest. The first column shows the segmentation inferred by model 1 and bounded to the aortic root region (in red). The second column shows the 3D anatomy reconstructions together with either the annulus (green), STJ (blue) or sinuses plane (yellow). The third column shows an interpolation of the image onto the plane identified by either the annulus, STJ or sinuses planes. Segmentation contours are show in green (annulus), blue (STJ) and yellow (sinuses), together with $D_{max}$ (dashed arrows, black) and $D_{min}$ (dashed arrows, white).}
	\label{fig7}
\end{figure}

For few cases, a large discrepancy was observed between the automatic and expert measurements. The largest differences in annulus diameters were observed for patient 057 and were equal to $\Delta D_{max}=-4.9 mm$ and $\Delta D_{min}=-5.7 mm$, where $\Delta$ is defined as (expert minus algorithm). However, a detailed analysis of case 057 showed that AR segmentation was precise; thus, discrepancies are not due to errors in automated segmentation (Figure \ref{fig7}, top row). Concerning STJ measurements, the largest discrepancies were obtained for patient 029 ($\Delta D_{mean}= 4.4 mm$); these were likely due to the presence of calcifications around the aortic bulb (Figure \ref{fig7}, mid row), possibly introducing larger uncertainty in STJ measurements. The largest difference in $D_{mean}$ for the sinuses of Valsalva was obtained for patient 140 ($\Delta D_{mean}= -4.3 mm$). In this case, the high curvature of the aortic bulb surface could cause larger discrepancies between the two methods (Figure \ref{fig7}, bottom row).

\section{Discussion}
In this work, we presented a fully automatic pipeline for aortic root morphological analysis and TAVI pre-procedural planning support from 3D CT. Our system provides 3D segmentations of the anatomical structures of interest without requiring any human supervision and extracts quantitative morphological parameters of the aortic root with good accuracy as compared to manual annotations performed by a clinical expert. \\
To obtain a quantitative morphometric assessment of the aortic root, our approach relies on two CNNs trained to perform automatic 3D segmentation of the ascending aorta including the LVOT (model 1), and of the aortic annulus and STJ (model 2). The CNN trained to segment the aorta and LVOT (model 1) proved able to accurately delineate the anatomical region of interest in all 31 validation cases from dataset A and all 178 test cases in dataset B. For the validation cases, the average Dice score and MSDs were comparable to state-of-the-art approaches focusing on similar anatomical districts, with values of 0.93 [0.82, 0.98] and 1.10 [0.22, 5.99] mm, respectively. In a previous paper by our group dealing with automatic segmentation of the thoracic aorta from CT, a mean Dice score of 0.954 was reported \cite{saitta2022deep}. Elattar et al. \cite{elattar2014automatic} adopted an aortic root segmentation approach based on normalized cuts and achieved Dice scores of 0.95 [0.85, 0.98] and MSDs of 0.74 $\pm$ 0.39 mm. In our study, model 1 was trained on ground truth segmentations encompassing different portions of the left ventricle due to CT scan field of view or inter-operator variability. This led to label map predictions with variable extents (as it is visible in Figure \ref{fig5}), entailing slightly worse accuracy with respect to other similar studies. This inconsistency was irrelevant to the downstream tasks of our pipeline and did not invalidate the overall effectiveness of our approach, which only requires the LVOT to be segmented correctly. In addition, when considering only the aortic root region mean values of Dice score and MSD of 0.96 and 0.14 mm were found. These errors are in the range of twice the average pixel spacing of 0.56 mm of our image dataset. 
Concerning the identification of the aortic annulus and STJ, our approach (model 2) differs from previously published techniques since it derives anatomical landmarks from image segmentation and differential geometry. Elattar et al. \cite{elattar2016automatic} based their methodology on extracting the aortic root centerline, computing the Gaussian curvature and an harmonic decomposition of the aortic shape. Their automatic measurements were in excellent agreement with manual ones, with differences in annulus radii of 0.24 $\pm$ 0.70 mm and 0.37 $\pm$ 0.82 mm for two observers, respectively, and differences in annulus angles of 6.86 $\pm$ 5.39\textdegree and 6.34 $\pm$ 4.00\textdegree. We obtained slightly larger differences for annulus mean radii (0.34 $\pm$ 0.78 mm), lower differences for annulus area-derived radii (0.006 $\pm$ 0.16 mm), and similar discrepancies for annulus angles (-3.07 $\pm$ 7.15\textdegree). However, \cite{elattar2016automatic} still represents a proof of concept, lacking generality, since automatic measurements are validated against a limited set of 40 patients. In our experience, Gaussian curvature-based criteria for landmark detection would perform suboptimally or fail for anatomies with less pronounced bulb curvature. Atlas-based methods could be an alternative, but could still be limited in their capability to generalize to very heterogeneous anatomies unless the reference atlas is heterogeneous enough \cite{lalys2019automatic}. The significantly larger number of test patients included in our study allowed us to devise robust solutions that proved able to deal with great anatomical variability among subjects. To date, with 178 patients, our study presents the most comprehensive validation of an automatic aortic root assessment system. 
Another type of landmark detection approach implies training a neural network to directly localize key anatomical points. Astudillo et al. \cite{astudillo2020automatic} adopt this kind of strategy using a DenseVNet architecture to detect coronary ostia position and height with respect to the annulus, validating their accuracy against manual annotations on 100 patients. Compared to manual annotations, they report mean differences of 0.54 mm and -0.16 mm in left and right coronary heights, and of 1.4 mm in 3D Euclidean distances of coronary ostia. However, when utilizing a landmark-based method for anatomical measurement quantification, an error in identifying even one landmark can result in significant errors in more complex anatomical metrics. In contrast, in our approach we did not rely on direct neural network-based landmark detection, but rather on a series of geometric computations for robust identification of the aortic annulus and STJ. In particular, the annulus and STJ plane refinement step, enabled our system to cope with segmentation errors, requiring only a rough localization of the two anatomical regions. This makes the results of our pipeline more dependent on the accuracy of aortic root and LVOT segmentation, which are more clearly defined and visible from CT scans, and thus pose an easier challenge to deep learning segmentation systems.
Given the large number of cases on which we tested our automatic measurement method, it is reasonable to expect some large discrepancies vs. manual measurements. In the critical cases reported in Figure \ref{fig7}, the automatic segmentation algorithms were able to segment the anatomical structures with good precision. However, some anatomical features such as high curvature of the aortic bulb, and presence of abundant calcium deposits could cause differences in how diameters are computed by the two approaches. Being able to automatically segment calcium, our system potentially enables the end-user to choose whether to include or exclude calcifications in the computed measurements.
Overall, our approach is significantly faster than commercially available semi-automated tools \cite{delgado2011automated}, requiring less than 45 seconds to run from CT to measurements on a GPU-accelerated workstation, therefore it shortens the processing time, making it compatible with the clinical routine, and applicable to large population studies. On average, the time required by the user for a full aortic assessment using commercially available software is around 30 minutes.

\section{Conclusions}
We have presented a pipeline for automating quantification of complex morphological biomarkers that are relevant for preprocedural planning of TAVI from CT images, providing an unprecedented validation of our approach against 178 patients, the largest to date. Our proposed method demonstrates a quick, accurate, and consistent assessment of aortic root anatomy from CT data, yielding 3D segmentations of the aortic root (mean Dice score 0.96). Incorporating this deep learning-based tool into the preoperative planning routine in TAVI environments could potentially lead to time and cost savings, as well as improved accuracy.

\clearpage
\bibliographystyle{elsarticle-num} 
\bibliography{biblio}

\end{document}